\documentclass[%
 twocolumn,
 superscriptaddress,
 preprintnumbers,
 nofootinbib,
 amsmath,amssymb,
 aps,
 prl
 ]{revtex4-2}
\usepackage{slashed,bm}
\usepackage{url}
\usepackage[colorlinks = true,
            linkcolor = blue,
            urlcolor  = blue,
            citecolor = blue,
            anchorcolor = blue]{hyperref}

\newcommand{\sign}{\mathop{\mathrm{sign}}\nolimits}
\newcommand{\tr}{\mathop{\mathrm{tr}}\nolimits}
\newcommand{\NP}{\slashed{\mathcal{P}}}
\newcommand{\BS}{\mathbb{S}}
\newcommand\HS{\mathrm{S}}

\def\nf{n_f}
\def\D#1{D_#1}

\def\frct#1#2{\mbox{\small{$\displaystyle\frac{#1}{#2}$}}}

\allowdisplaybreaks

\begin{document}

\preprint{PUBDB-2026-01395, LTH 1415}

\title{Four-Loop Gluon Anomalous Dimension of General Lorentz Spin: Transcendental Part}

\author{B.A.~Kniehl}
\email{kniehl@desy.de}
\affiliation{II. Institute for Theoretical Physics, Hamburg University, 
 D-22761 Hamburg, Germany} 
\author{S.-O.~Moch}
\email{sven-olaf.moch@desy.de}
\affiliation{II. Institute for Theoretical Physics, Hamburg University, 
 D-22761 Hamburg, Germany} 
\author{V.N.~Velizhanin}
\email{vitaly.velizhanin@desy.de}
\affiliation{II. Institute for Theoretical Physics, Hamburg University, 
 D-22761 Hamburg, Germany} 
\author{A.~Vogt}
\email{andreas.vogt@liverpool.ac.uk}
\affiliation{II. Institute for Theoretical Physics, Hamburg University, 
 D-22761 Hamburg, Germany} 
\affiliation{Department of Mathematical Sciences, University of Liverpool, 
 Liverpool L69 3BX, United Kingdom}


\begin{abstract}
  We consider the anomalous dimension $\gamma_{gg}^{(3)}(N)$ of the twist-two gluon operator of arbitrary Lorentz spin $N$ in the quark flavor singlet sector of a general gauge theory at four loops and construct its contribution proportional to $\zeta(3)$ in analytic form by applying the Lenstra-Lenstra-Lov\'{a}sz algorithm to the available low-$N$ moments.
We exploit generalized Gribov--Liptov reciprocity, establish new self-tuning relations for the anomalous dimension matrix of the singlet sector, and inject information from $\mathcal{N}=1,4$ supersymmetric Yang--Mills theories.
We also present the contribution to the rational part of $\gamma_{gg}^{(3)}(N)$ with color factor $C_F^2n_f^2$.
Exact contributions to the four-loop splitting function $P_{gg}^{(3)}(x)$ hence resulting via inverse Mellin transformation help us to reduce theoretical uncertainties in scaling violations of parton distribution functions in QCD.
\end{abstract}

\pacs{}

\maketitle

%

The exploitation of the physics potentials of future high-precision measurements at the CERN Large Hadron Collider, the BNL Electron-Ion Collider, and other hadron-beam facilities critically depends on our ability to deepen our understanding of the proton's parton density functions (PDFs), which enter every cross section prediction in the parton model \cite{Bjorken:1969ja} of quantum chromodynamics (QCD) \cite{Fritzsch:1973pi}.
While PDFs are genuinely nonperturbative objects, their scaling violations, encoded in Dokshitzer--Gribov--Lipatov--Altarelli--Parisi (DGLAP) \cite{Gribov:1972ri,Gribov:1972rt,Altarelli:1977zs,Dokshitzer:1977sg} evolution, are amenable to perturbation theory in the strong-coupling constant $\alpha_s$.
This and PDF universality, both consequences of the factorization theorem \cite{Collins:1989gx}, provide the theoretical foundation both for determinations of PDFs from global fits to experimental data and their predictive power for hadron-induced processes.
At the same time, this allows for high-precision determinations of $\alpha_s$ \cite{Alekhin:2025qdj} and heavy-quark masses \cite{Alekhin:2010sv,Alekhin:2012un}.

The DGLAP evolution kernels consist of splitting functions, $P_{ij}(x)$, which, in a way, measure the probability of parton (gluon or quasi-massless quark) $j$ to inclusively branch into collinear parton $i$ carrying fraction $x$ of $j$'s momentum.
The perturbative expansion
\begin{equation}
P_{ij}(x)=\sum_{n=0}^{\infty}a_s^{n+1}P_{ij}^{(n)}(x)\,,
\end{equation}
with $a_s=\alpha_s(\mu)/(4\pi)$, is a topic of old vintage, dating back to the early 1970s, when the leading order (LO) was explored \cite{Gribov:1972ri,Gribov:1972rt,Altarelli:1977zs,Dokshitzer:1977sg,Gross:1973ju,Gross:1974cs}, and has enjoyed top priority on the agenda of the world-wide particle physics community ever since.
Full analytic expressions are available at next-to-LO (NLO) \cite{Floratos:1977au,GonzalezArroyo:1979df,Gonzalez-Arroyo:1979kjx,Curci:1980uw,Furmanski:1980cm} and N${}^2$LO \cite{Larin:1993vu,Larin:1996wd,Moch:2004pa,Vogt:2004mw,Ablinger:2014nga,Blumlein:2021enk,Blumlein:2022gpp,Gehrmann:2023ksf}, while only partial results are yet known at N${}^3$LO \cite{Gracey:1994nn,Velizhanin:2011es,Velizhanin:2014fua,Davies:2016jie,Moch:2017uml,Davies:2017hyl,Moch:2018wjh,Das:2020adl,Moch:2021qrk,Falcioni:2023luc,Gehrmann:2023cqm,Falcioni:2023vqq,Falcioni:2023tzp,Moch:2023tdj,Gehrmann:2023iah,Falcioni:2024xyt,Falcioni:2024xav,Falcioni:2024qpd,Kniehl:2025jfs,Kniehl:2025ttz,Falcioni:2025hfz,Kniehl:2026axe,Gehrmann:2026qbl}.
Modern evaluations exploit the fact that the Mellin transforms
\begin{equation}
  \gamma_{ij}(N)=-\int_0^1 dx\ x^{N-1} P_{ij}(x)\,,  
\label{eq:mel}
\end{equation}
are anomalous dimensions of local composite operators of twist two and Lorentz spin $N$, constructed from quark fields, gluon field strength tensors, and a definite number of covariant derivatives, namely $N-1$ for quark operators and $N-2$ for gluon operators \cite{Dokshitzer:1977sg}.
Specifically, one renormalizes off-shell matrix elements of such operators in the $\overline{\mathrm{MS}}$ scheme.
At N$^3$LO, one must deal with four-loop Feynman diagrams, whose complexity rapidly grows with increasing $N$ until the limits of available computer algebra systems and computing resources are reached.
In turn, from finite sets of Mellin moments, only approximate results for $P_{ij}$ can be gained \cite{Larin:1996wd,Moch:2017uml,Moch:2023tdj,Falcioni:2024xyt,Falcioni:2025hfz}, which generally become insufficient in the high-energy limit, at very low $x$ values (see, {\it e.g.}, Figs.~5 and 6 in Ref.~\cite{Moch:2017uml}), as probed by the LHC.
This strongly motivates us to recover all-$N$ results for $\gamma_{ij}^{(3)}$ as much as possible.

Given the PDFs of $f$-flavored quarks and antiquarks and the gluon, $q_f$, $\bar{q}_f$, and $g$, the quark sector is organized in terms of flavor asymmetries $q_{\mathrm{ns},ff^\prime}^{\pm}=(q_f\pm \bar{q}_f)-(q_{f^\prime}\pm \bar{q}_{f^\prime})$, flavor non-singlet valence $q_{\mathrm{ns}}^{\mathrm{v}}=\sum_{f=1}^{n_f}(q_f-\bar{q}_f)$, and flavor singlet $q_{\mathrm{s}}=\sum_{f=1}^{n_f}(q_f+\bar{q}_f)$, where $n_f$ is the number of active quark flavors.
Then the DGLAP evolution decouples for $q_{\mathrm{ns},ff^\prime}^{\pm}$ and $q_{\mathrm{ns}}^{\mathrm{v}}$, leaving a $2\times2$ matrix equation for $q_{\mathrm{s}}$ and $g$.
This involves seven distinct splitting functions, $P_{\mathrm{ns}}^{\pm}$, $P_{\mathrm{ns}}^{\mathrm{v}}$, $P_{qq}$, $P_{qg}$, $P_{gq}$, $P_{gg}$.
Furthermore, one decomposes $P_{\mathrm{ns}}^{\mathrm{v}}=P_{\mathrm{ns}}^-+P_{\mathrm{s}}$ and $P_{qq}=P_{\mathrm{ns}}^++P_{\mathrm{ps}}$, involving quark sea and pure singlet.
This very structure is mapped to Mellin space via Eq.~\eqref{eq:mel}.

In this letter, we focus attention on $\gamma_{gg}^{(3)}$, at N$^3$LO.
This is known to break down in rational and transcendental parts as \cite{Moch:2021qrk,Moch:2023tdj,Falcioni:2024qpd,Falcioni:2025hfz}
\begin{equation}
\gamma_{gg}^{(3)}=
\gamma_{gg,\mathrm{rat}}^{(3)}
+\zeta_3\gamma_{gg,\zeta_3}^{(3)}
+\zeta_4\gamma_{gg,\zeta_4}^{(3)}
+\zeta_5\gamma_{gg,\zeta_5}^{(3)}\,,
\label{eq:zeta3}
\end{equation}
where $\zeta_k=\zeta(k)$ is Riemann's zeta function.
A term proportional to $\zeta_2=\pi^2/6$ is prohibited by the no-$\pi^2$ theorem \cite{Davies:2017hyl,Jamin:2017mul,Baikov:2018wgs} of the $\overline{\mathrm{MS}}$ scheme.
Presently, $\gamma_{gg,\zeta_5}^{(3)}$ \cite{Moch:2018wjh,Falcioni:2024qpd}, $\gamma_{gg,\zeta_4}^{(3)}$ \cite{Davies:2017hyl}, the $n_f^3$ \cite{Davies:2016jie} and $n_f^2$ \cite{Falcioni:2024qpd} terms of $\gamma_{gg,\zeta_3}^{(3)}$, and the $n_f^3$ term of $\gamma_{gg,\mathrm{rat}}^{(3)}$ \cite{Davies:2016jie} are known for all values of $N$, while the residual terms, namely the $n_f^{1,0}$ terms of $\gamma_{gg,\zeta_3}^{(3)}$ and the $n_f^{2,1,0}$ terms $\gamma_{gg,\mathrm{rat}}^{(3)}$, are only available for $N=2,4,\ldots,22$ \cite{Moch:2021qrk
,Moch:2023tdj
,Falcioni:2024qpd
,Falcioni:2025hfz
}.
The results for odd values of $N$ are inaccessible via the operator product expansion.
In the following, we analytically reconstruct the all-$N$ result for $\gamma_{gg,\zeta_3}^{(3)}$ utilizing advanced techniques of number theory in combination with an understanding of the appearing classes of special functions, conjectured theorems, and educated guesses.
This completes our knowledge of the transcendental part of $\gamma_{gg}^{(3)}$ for arbitrary value of $N$.
Recently, the same has been done for $\gamma_{\mathrm{ns}}^{(3)\pm}$, $\gamma_{\mathrm{s}}^{(3)}$ \cite{Kniehl:2025jfs,Kniehl:2026axe,Gehrmann:2026qbl}, $\gamma_{\mathrm{ps}}^{(3)}$ \cite{Falcioni:2025hfz}, and partly also for $\gamma_{qg}^{(3)}$ and $\gamma_{gq}^{(3)}$, up to $n_f^1$ and $n_f^0$ terms, respectively \cite{Falcioni:2025hfz}.

Detailed inspection of the known all-$N$ expressions for $\gamma_{ij}^{(n)}$, at $n+1$ loops, and parts thereof reveals that the special functions appearing therein are exhausted by the nested harmonic sums \cite{Vermaseren:1998uu},
\begin{equation} 
S_{m_1,\ldots,m_M}(N)=\sum^{N}_{i=1} \frct{[\sign(m_1)]^{i}}{i^{\vert m_1\vert}}
\,S_{m_2,\ldots,m_M}(i)\,,
\label{eq:hs}
\end{equation}
where $S(N)=1$ and $m_j\in\mathbb{Z}\backslash\{-1,0\}$, with weight (transcendentality) $w=\sum_{j=1}^M|m_j|\le2n+1$.
At weight $w$, there are $[(1-\sqrt{2})^w+(1+\sqrt{2})^w]/2$ of them.
In addition, their counterparts with shifted arguments $N\pm1$ and $N\pm2$ may occur.
These may be rewritten as linear combinations of $D_1$, $D_2$, $\eta$, and $\nu$, having $w=1$, their powers, and products of these terms with nested harmonic sums of argument $N$, where
\begin{equation}
\eta=D_0-D_1\,,\qquad \nu=D_{-1}-D_2\,,\qquad D_k=\frac{1}{N+k}\,.
\label{eq:etanu}
\end{equation}
Notice that $\zeta_k$ has $w=k$ and its appearance as an overall factor correspondingly reduces the maximum weight of the function basis used for the ansatz of analytic reconstruction.
Fortunately, the coefficients in the ansatz empirically come as relatively harmless fractions, made of integers of moderate size modulo some powers of 2 and 3.
Therefore, we may attempt to determine them using the Lenstra-Lenstra-Lov\'{a}sz (LLL) algorithm \cite{Lenstra82factoringpolynomials} as implemented in the program package \texttt{fplll} \cite{fplll}, which allows us to solve systems of 
linear equations with far more unknowns than equations.
A reliable success indicator is that several coefficients vanish and that the remaining ones are simple fractions.
A typical example of how this comes about is presented in the Supplemental Material~\ref{supplement:1}.
This method for reconstructing all-$N$ expressions of anomalous dimensions from a finite set of Mellin moments was pioneered in $\mathcal{N}=4$ supersymmetric Yang--Mills theory (SYM) \cite{Velizhanin:2010cm} and subsequently applied in ${\mathcal N}=4$ \cite{Velizhanin:2013vla,Marboe:2014sya,Marboe:2016igj,Kniehl:2020rip,Kniehl:2021ysp,Velizhanin:2021bdh,Kniehl:2023bbk,Kniehl:2024tvd} and $\mathcal{N}=2$ SYM \cite{Kniehl:2023bbk} and also in QCD \cite{Velizhanin:2012nm,Davies:2016jie,Moch:2017uml,Kniehl:2025jfs,Kniehl:2025ttz,Falcioni:2025hfz,Kniehl:2026axe}.
Prior to applying the LLL algorithm \cite{Lenstra82factoringpolynomials,fplll}, we may fix the coefficients of $\eta^{4,3,2}$ and $\nu^4$ using the Chinese remainder theorem \cite{china} as originally done in Ref.~\cite{Moch:2014sna}.
According to the na\"{\i}ve weight counting rules mentioned above, the coefficient of $\zeta(3)$ has $w=4$, so that there are hundreds of nested harmonic sums, which is far beyond what the LLL algorithm can manage with just eleven $\gamma_{gg,\zeta_3}^{(3)}$ inputs.

This situation would be hopeless, if it were not for Gribov--Lipatov reciprocity, originally observed at LO as $P_{ii}^{(0)}(x)=-xP_{ii}^{(0)}(1/x)$ or, equivalently, as the quasi-invariance of $\gamma_{ii}^{(0)}(N)$ under the mapping $N\to-1-N$ \cite{Gribov:1972rt}.
This incorporates a similar relation between time-like and space-like structure functions previously established by Drell, Levy, and Yan \cite{Drell:1969jm,Blumlein:2000wh}.
In the non-singlet sector, the $N\leftrightarrow-1-N$ symmetry was recovered also beyond LO, albeit in the alleviated sense that it no longer holds for the full anomalous dimension $\gamma(N)$, but only for the so-called reciprocity-respecting part $\mathcal{P}(N)$ of its representation in terms of nested harmonic sums in the large-$N$ limit, which is known to satisfy the self-tuning relation \cite{Dokshitzer:2005bf,Dokshitzer:2006nm,Basso:2006nk},
\begin{equation}
\gamma(N)=\mathcal{P}(N-\gamma(N)-\beta(a_s)/a_s)\,,
\label{eq:self}
\end{equation}
where
\begin{equation}
\mu^2\frct{\mathrm{d}a_s}{\mathrm{d}\mu^2}=
\beta(a_s)=-\sum_{n=0}^\infty b_n\,a_s^{n+2}\,,
\end{equation}
is the $\beta$ function of QCD \cite{Gross:1973id,Politzer:1973fx,Jones:1974mm,Caswell:1974gg,Tarasov:1980au,Larin:1993tp,vanRitbergen:1997va,Chetyrkin:2004mf,Czakon:2004bu,Luthe:2016ima,Baikov:2016tgj,Herzog:2017ohr,Luthe:2017ttg}.
This entails the decomposition $\mathcal{P}(N)=\gamma(N)+\NP(N)$, where the coefficients $\NP^{(n)}(N)$ of the non-reciprocity-respecting remainder,
\begin{equation}
  \NP(N)=\sum_{n=1}^\infty a_s^{n+1}\NP^{(n)}(N)\,,
\end{equation}
are expressed entirely in terms of $\gamma^{(k)}$, its derivatives with respect to $N$, and $b_k$ with $k<n$.
Explicit expressions for $\NP^{(n)}$ with $n=1,2,3$ may be found in Refs.~\cite{Kniehl:2025jfs,Kniehl:2026axe}.
In other words, $\NP^{(3)}$ is fully determined by known results from N${}^2$LO and below.
The salient point is that the function space thus collapses to the much smaller reciprocity-respecting subspace of binomial harmonic sums \cite{Vermaseren:1998uu},
\begin{equation}
\BS_{m_1,\ldots,m_M}\!(N)\!=\!(-1)^N\!\sum_{i=1}^{N}\!(-1)^i\!\binom{N}{i}\!\!\binom{N+i}{i}\!S_{m_1,...,m_M}(i)\,,
\end{equation}
where $m_j\in\mathbb{N}$.
In fact, there are only $2^{w-1}$ such sums with weight $w$.
Similarly to the case of nested harmonic sums, we have to allow for the binomial harmonic sums to be dressed by powers of $\eta$ and $\nu$, which are both reciprocity respecting.

In the singlet sector, Eq.~\eqref{eq:self} does not hold for the individual entries of the matrix
\begin{equation}
\hat{\gamma}=\left( \begin{array}{cc}
         \gamma_{qq} & \gamma_{qg} \\
         \gamma_{gq} & \gamma_{gg} \end{array} \right)\,,
\label{eq:matrix}
\end{equation}
where $\gamma_{qq}=\gamma_{\mathrm{ns}}^++\gamma_{\mathrm{ps}}$.
However, detailed inspection reveals that the eigenvalues, $\gamma_\pm=\tr\hat{\gamma}/2\pm[(\tr\hat{\gamma})^2/4-\det\hat{\gamma}]^{1/2}$, do satisfy Eq.~\eqref{eq:self}.
This follows from Ref.~\cite{Basso:2006nk} and further studies of large-spin systematics~\cite{Alday:2015eya}.
A similar observation was made at N${}^2$LO in Ref.~\cite{Chen:2020uvt}, mixing spacelike and timelike quantities, however.  
Our formulation has the advantage of being purely spacelike.
More details will be presented in Ref.~\cite{Moch-Shumilov-to-appear}.
Because of the linearity of the decomposition into reciprocity-respecting and non-reciprocity-respecting parts, also $\gamma_++\gamma_-=\tr\hat{\gamma}=\gamma_{qq}+\gamma_{gg}$ satisfies Eq.~\eqref{eq:self}.
Exploiting the knowledge of $\gamma_{\mathrm{ns},\zeta_3}^{(3)+}$ \cite{Kniehl:2025jfs,Kniehl:2026axe,Gehrmann:2026qbl} and $\gamma_{\mathrm{ps},\zeta_3}^{(3)}$ \cite{Falcioni:2025hfz} for all $N$, we may thus analytically reconstruct $\tr\hat{\gamma}_{\zeta_3}^{(3)}$ from the known Mellin moments of $\gamma_{gg,\zeta_3}^{(3)}$ \cite{Moch:2021qrk,Moch:2023tdj,Falcioni:2024qpd,Falcioni:2025hfz} and so obtain $\gamma_{gg,\zeta_3}^{(3)}=\tr\hat{\gamma}_{\zeta_3}^{(3)}-\gamma_{\mathrm{ns},\zeta_3}^{(3)+}-\gamma_{\mathrm{ps},\zeta_3}^{(3)}$ for all $N$.

The sought-after $n_f^1$ and $n_f^0$ terms of $\gamma_{gg,\zeta_3}^{(3)}$ come with color factors $n_fC_F^3$, $n_fC_AC_F^2$, $n_fC_A^2C_F$, $n_fC_A^3$, $n_fd_{RA}^{(4)}/n_a$, $C_A^4$, and $d_{AA}^{(4)}/n_a$ \cite{Moch:2021qrk,Moch:2023tdj,Falcioni:2024qpd,Falcioni:2025hfz}, where $C_F=(n_c^2-1)/(2n_c)$, $C_A=n_c$, $d_{RA}^{(4)}=d_F^{abcd}d_A^{abcd}/n_a=n_c(n_c^2+6)/48$ and $d_{AA}^{(4)}/n_a=d_A^{abcd}d_A^{abcd}/n_a=n_c^2(n_c^2+36)/24$ with $n_a=n_c^2-1$ for color gauge group SU($n_c$).
By the conjectured maximal-transcendentality principle \cite{Kotikov:2002ab,Kotikov:2006ts}, the $w=4$ terms in the coefficients of $C_A^4$ and $d_{AA}^{(4)}/n_a$ may be gleaned respectively from the planar \cite{Beisert:2006ez,Bajnok:2008qj} and nonplanar \cite{Kniehl:2020rip,Kniehl:2021ysp,Kniehl:2024tvd} universal anomalous dimensions of the twist-two operators with general Lorentz spin $N$ at N${}^3$LO in $\mathcal{N}=4$ SYM, which serves as welcome input.
Since the quartic color factors $n_fd_{RA}^{(4)}/n_a$ and $d_{AA}^{(4)}/n_a$ appear for the first time at N${}^3$LO, their coefficients are reciprocity respecting and may be reconstructed directly.
In terms of binomial harmonic sums, we find
\begin{widetext}
\begin{eqnarray}
&&  \gamma_{gg,\zeta_3}^{(3)}\Big|_{\mathrm{quar}}=
%
%
\frct{128}{3}\*\frct{d_{RA}^{(4)}}{n_a}\*n_f\*\Big[  
\frct{103}{15}
+\frct{2908}{45}\*\nu
-291\*\eta 
-387\*\eta^2
+\BS_{1}\*\Big(\frct{52}{3}\*\nu +5 + 14\*\eta - 46\*\eta ^2 - 84\*\eta ^3\Big)
-\BS_{2}\*\Big(\frct{514}{15}+\frct{618}{5}\*\eta
- 4\*\nu 
\nonumber\\
&&{}  
- 132\*\eta ^2\Big)
+\BS_{1,1}\*\big(9\*\eta - 16\*\nu + 198\*\eta ^2\big)
+10\*\BS_{1,2}
+\BS_{2,1}\*\big(23-18\*\eta\big)
+\BS_{3}\*\big(36\*\eta-36\big)
+(N+1)\*N\*\Big(-\frct{2}{15}\*\BS_{2}\
+6\*\BS_{3} -3\*\BS_{2,1}\Big)
\Big]
\nonumber\\
&&{}
+\frct{64}{3}\*\frct{d_{AA}^{(4)}}{n_a}\*
  \Big[  
 \frct{331}{15} 
+ \frct{429}{2}\*\eta
- \frct{3824}{45}\*\nu
+ 180\*\eta^2 
+ \BS_{1} \*\Big(-5 + 1352\*\eta - \frct{1328}{3}\*\nu + 722\*\eta^2 + 144\*\eta^3 + 192\*\nu^2\Big)
+ 2\*\BS_{2}\*\Big(\frct{1126}{15}+\frct{1127}{5}\*\eta 
\nonumber\\ 
&&{}
- 19\*\nu + 48\*\nu^2 - 12\*\eta^2 \Big)
- 8\*\BS_{1, 1}\*(63\*\eta + 54\*\eta^2 - 25\*\nu +12\*\nu^2)
+ 9\*\BS_{3}\*(13 - 28\*\eta + 12\*\nu)
- \BS_{1, 2}\*(55 - 48\*\eta + 48\*\nu)
- 2\*\BS_{2, 1}\*(43 
\nonumber\\ 
&&{}  
- 66\*\eta + 30\*\nu)
+ 48\*\BS_{4} 
- 54\*\BS_{1, 3} 
- 18\*\BS_{2, 2} 
- 24\*\BS_{3, 1} 
+ 24\*\BS_{1, 1, 2} 
+ 30\*\BS_{1, 2, 1} 
- 6\*\BS_{2, 1, 1} 
+ (N+1)\*N\*\Big(\frct{1}{15}\*\BS_{2}\
-6\*\BS_{3} +3\*\BS_{2,1}\Big) 
\Big]
\,,\qquad
\label{eq:quartic}  
\end{eqnarray}
\end{widetext}
where $\eta$ and $\nu$ are defined in Eq.~\eqref{eq:etanu} and we have omitted the arguments $N$ of the binomial harmonic sums.
The coefficients of the quadratic color factors require the full procedure described above.
To enable cancellations, we rewrite each binomial harmonic sum in the expression thus obtained in terms of nested harmonic sums.
We have
\begin{widetext}
\begin{eqnarray}
&&\gamma_{gg,\zeta_3}^{(3)}\Big|_{\mathrm{quad}}=
%
%
\frct{32}{3}\*C_F^3\*n_f\*
\Big[
24\*\D1
-24\*\D2
+231\*\D1^2
+294\*\D1^3
+144\*\D1^4
-\frct{897}{4}\*\eta
-16\*\nu
-177\*\eta^2
-138\*\eta^3
-114\*\eta^4
+6\*(
6\*\D1
\nonumber\\
&&{}
-6\*\D2
-2\*\D1^2
+8\*\D1^3
-71\*\eta
+20\*\nu
-78\*\eta^2
-56\*\eta^3
)\*\HS_{1}
-4\*(15\*\eta
-4\*\nu
+18\*\eta^2
)\*\HS_{2}
+4\*(33\*\eta
-16\*\nu
+90\*\eta^2
)\*\HS_{-2}
+12\*(13\*\eta
\nonumber\\
&&{}-4\*\nu
+14\*\eta^2
)\*\HS_{1,1}
-\frct{2}{3}\*\delta_{N,2}
\Big]
+\frct{32}{9}\*C_A\*C_F^2\*n_f\*\Big\{
-\frct{11}{2}
+\frct{1161}{2}\*(\D1-\D2)
-2817\*\D1^2
-1476\*\D1^3
-1836\*\D1^4
+148\*\D2^2
\nonumber\\
&&{}
+120\*\D2^3
+176\*\nu^2
+\frct{6879}{2}\*\eta
-\frct{1597}{3}\*\nu
+\frct{3771}{2}\*\eta^2
+2376\*\eta^3
+1224\*\eta^4
+3\*\Big[\frct{37}{2}
-30\*(\D1-\D2)
+123\*\D1^2
+12\*\D2^2
+132\*\D1^3
\nonumber\\
&&{}
+1065\*\eta
-366\*\nu
+1656\*\eta^2
-8\*\nu^2
+930\*\eta^3
\Big]\*\HS_{1}
+3\*(231\*\eta
-68\*\nu
+330\*\eta^2
)\*\HS_{2}
-24\*(24\*\eta
-13\*\nu
+81\*\eta^2
)\*\HS_{-2}
-48\*(30\*\eta
-10\*\nu
\nonumber\\
&&{}
+39\*\eta^2
)\*\HS_{1,1}
+\frct{7}{3}\*\delta_{N,2}
\Big\}
+\frct{32}{3}\*C_A^2\*C_F\*n_f\*\Big\{\frct{2}{3}
-\frct{471}{2}\*(\D1-\D2)
+\frct{2019}{2}\*\D1^2
-\frct{514}{3}\*\D2^2
+201\*\D1^3
-136\*\D2^3
+684\*\D1^4
\nonumber\\
&&{}
-\frct{2530}{3}\*\eta
+\frct{1303}{18}\*\nu
-\frct{1325}{3}\*\eta^2
-\frct{256}{3}\*\nu^2
-892\*\eta^3
-40\*\nu^3
-420\*\eta^4
+\Big[\frct{241}{6}
-6\*(\D1-\D2)
-135\*\D1^2
-84\*\D2^2
-324\*\D1^3
\nonumber\\
&&{}
-552\*\eta
+\frct{773}{3}\*\nu
-1470\*\eta^2
+112\*\nu^2
-726\*\eta^3
\Big]\*\HS_{1}
-(
309\*\eta
-124\*\nu
+426\*\eta^2
)\*\HS_{2}
+2\*(3
-3\*\eta
-8\*\nu
+210\*\eta^2
)\*\HS_{-2}
+8\*(60\*\eta
\nonumber\\
&&{}
-23\*\nu
+90\*\eta^2
)\*\HS_{1,1}
+8\*\HS_{3}
+8\*\HS_{-3}
-24\*\HS_{1,2}
-24\*\HS_{1,-2}
-24\*\HS_{2,1}
+8\*\HS_{-2,1}
-\frct{2}{9}\*\delta_{N,2}
\Big\}
+\frct{32}{3}\*C_A^3\*n_f\*\Big[
\frct{2197}{360}
-294\*\D1^2
+122\*\D2^2
\nonumber\\
&&{}
-24\*\D1^3
+96\*\D2^3
-216\*\D1^4
+\frct{683}{12}\*\eta
+\frct{6734}{135}\*\nu
+\frct{2431}{12}\*\eta^2
+\frct{194}{3}\*\nu^2
+377\*\eta^3
+20\*\nu^3
+144\*\eta^4
+\frct{1}{18}\*(
-1173
+432\*\D1^2
\nonumber\\
&&{}
+1296\*\D2^2
+2592\*\D1^3
+4989\*\eta
-1696\*\nu
+8913\*\eta^2
-1440\*\nu^2
+4194\*\eta^3
)\*\HS_{1}
+\frct{5}{3}\*(99\*\eta
-40\*\nu
+54\*\eta^2
)\*\HS_{2}
-\frct{2}{3}\*(
315\*\eta
-92\*\nu
\nonumber\\
&&{}
+162\*\eta^2
)\*\HS_{1,1}
+\frct{1}{6}\*\Big(
\frct{128}{15}
+\frct{2436}{5}\*\eta
-116\*\nu
-1056\*\eta^2
\Big)\*\HS_{-2}
-\frct{1}{6}\*(
59
-18\*\eta)\*\HS_{3}
+\frct{1}{3}\*(31
-18\*\eta)\*\HS_{-3}
+24\*\HS_{1,2}
+\frct{46}{3}\*\HS_{1,-2}
\nonumber\\
&&{}
+24\*\HS_{2,1}
+12\*(
-3
+\eta)\*\HS_{-2,1}
+\frct{1}{2}\*N\*(N+1)\*\Big(
\frct{4}{45}\*\HS_{-2}
+\HS_{3}
-2\*\HS_{-3}
+4\*\HS_{-2,1}\Big)\Big]
+\frct{64}{3}\*C_A^4\*\Big[
-\frct{2621}{720}
-\frct{36013}{144}\*\eta
+\frct{101027}{1080}\*\nu
\nonumber\\
&&{}
-\frct{1133}{6}\*\eta^2
-\frct{207}{4}\*\nu^2
-97\*\eta^3
+19\*\nu^3
-12\*\eta^4
-24\*\nu^4
+\frct{1}{36}\*(437
-21021\*\eta
+5915\*\nu
-12174\*\eta^2
-2664\*\nu^2
-4320\*\eta^3
\nonumber\\
&&{}
+1728\*\nu^3)\*\HS_{1}
+\frct{1}{3}\*(5
-51\*\eta
+33\*\nu)\*\HS_{3}
+6\*\HS_{4}
+\frct{4}{3}\*(45\*\eta
+45\*\eta^2
-25\*\nu
+12\*\nu^2)\*(\HS_{2}-2\*\HS_{1,1})
-\frct{1}{18}\*\Big(
\frct{452}{5}
-\frct{1023}{5}\*\eta
+435\*\nu
\nonumber\\
&&{}
-612\*\eta^2
+72\*\nu^2
\Big)\*\HS_{-2}
-\frct{1}{3}\*(
131
-141\*\eta
+105\*\nu)\*\HS_{-3}
-18\*\HS_{-4}
-11\*\HS_{1,3}
+\frct{1}{6}\*(143
-96\*\eta
+96\*\nu)\*\HS_{1,-2}
+(127
-156\*\eta
\nonumber\\
&&{}
+108\*\nu)\*\frct{1}{2}\*\HS_{-2,1}
+35\*\HS_{1,-3}
+8\*\HS_{2,-2}
-5\*\HS_{3,1}
+4\*\HS_{-2,-2}
+16\*\HS_{-2,2}
+43\*\HS_{-3,1}
-16\*\HS_{1,1,-2}
-54\*\HS_{1,-2,1}
-32\*\HS_{-2,1,1}
\nonumber\\
&&{}
-N\*(N+1)\*\Big(
\frct{1}{90}\*\HS_{-2}
+\frct{1}{4}\*\HS_{3}
-\frct{1}{2}\*\HS_{-3}
+\HS_{-2,1}
\Big)
\Big]
\,,
\label{eq:quadratic}
\end{eqnarray}
\end{widetext}
where $D_k$ is defined in Eq.~\eqref{eq:etanu} and we have omitted the arguments $N$ of the nested harmonic sums.
The $\zeta_3\delta_{N,2}$ terms with $\{C_F^2,C_AC_F,C_A^2\}C_Fn_f$ in Eq.~\eqref{eq:quadratic} are familiar from the $C_AC_Fn_f^2$ contribution to $\gamma_{gg,\zeta_3}^{(3)}$ \cite{Falcioni:2024qpd} and similar contributions to $\gamma_{\mathrm{ps}}^{(3)}$ with $C_AC_Fn_f^2$ \cite{Gehrmann:2023cqm} and $\gamma_{gq}^{(3)}$ with $C_AC_Fn_f^2$ \cite{Falcioni:2023tzp}.
Such terms do not yet appear in $\gamma_{ij}^{(2)}$ \cite{Larin:1993vu,Larin:1996wd,Moch:2004pa,Vogt:2004mw,Ablinger:2014nga,Blumlein:2021enk,Blumlein:2022gpp,Gehrmann:2023ksf}, but were already observed at two loops in the coefficient functions of inclusive deep-inelastic scattering (DIS) by photon exchange \cite{vanNeerven:1991nn,Moch:1999eb}.
They emerge from $D_{-2}S_{-2}(N-2)$ for $N=2$ and allow us to predict that $\gamma_{gg,\mathrm{rat}}^{(3)}$ will contain the term $(64/81)(6C_F^2-7C_AC_F+2C_A^2)C_Fn_f\theta(N-4)D_{-2}S_{-2}(N-2)$.
The $\zeta_3N(N+1)\BS_2$ terms with $n_fd_{RA}^{(4)}$ and $d_{AA}^{(4)}$ in Eq.~\eqref{eq:quartic} and with $n_fC_A^3$ and $C_A^4$ in Eq.~\eqref{eq:quadratic} (notice that $S_{-2}=-\BS_2/2$) conspire with similar terms in $\gamma_{gg,\zeta_5}^{(3)}$ \cite{Moch:2018wjh,Falcioni:2024qpd} to produce the first two terms of the well-known factor
\begin{equation}
f(N) = 5 \zeta_5 - 2 \zeta_3 \BS_{2}(N) - \BS_{2, 1, 2}(N) + \BS_{3, 1, 1}(N)\,,
\label{eq:fn}
\end{equation}
which was first encountered in three-loop coefficient functions of inclusive DIS \cite{Vermaseren:2005qc,Moch:2008fj} and later recovered in wrapping corrections to twist-two operators at four loops in $\mathcal{N}=4$ SYM \cite{Bajnok:2008qj}.
The analogous observation was made in Ref.~\cite{Falcioni:2024qpd} for the $\{C_A^2,d_{FF}^{(4)}/n_a\}n_f^2N(N+1)\{\zeta_5,\zeta_3\BS_{2}\}$ terms of $\gamma_{gg}^{(3)}$, where $d_{RR}^{(4)}/n_a=d_R^{abcd}d_R^{abcd}/n_a=(n_c^4-6n_c^2+18)(n_c^2-1)/(96n_c^3)$.
We thus conjecture that $\gamma_{gg,\mathrm{rat}}^{(3)}$ will contain the term $c_1N(N+1)(- \BS_{2, 1, 2} + \BS_{3, 1, 1})$, where $c_1=-(8/45)[C_A^2(C_A-n_f)^2/3+4(d_{AA}^{(4)}-4d_{RA}^{(4)}n_f+4d_{RR}^{(4)}n_f^2)/n_a]$.
A similar conjecture for $\gamma_{\mathrm{ns},\mathrm{rat}}^{(3)\mathrm{s}}$ \cite{Kniehl:2026axe} has recently been confirmed in Ref.~\cite{Gehrmann:2026qbl}.
In this way, these $N(N+1)$ power terms will be tamed to comply with the leading large-$N$ behavior of $\gamma_{gg}^{(3)}$ being $\ln N$ times the cusp anomalous dimension \cite{Korchemsky:1988si} because $f\propto\eta$ in this limit.
By the same token, the power divergence of the previously unobserved term $c_2N(N+1)(2\BS_3-\BS_{2,1})$, with $c_2=8[C_A^3(C_A-n_f)/3-8(d_{AA}^{(4)}-2d_{RA}^{(4)}n_f)/n_a]$, in $\gamma_{gg,\zeta_3}^{(3)}$ must be canceled by an appropriate counterpart in $\gamma_{gg,\mathrm{rat}}^{(3)}$, where we have used $2\BS_3-\BS_{2,1}=2(2S_{-3}-S_3-4S_{-2,1})$.
We note that $\mathcal{N}=1$ SYM is devoid of such terms because $c_1=c_2=0$ for $C_A=n_f$ and $d_{AA}^{(4)}=2d_{RA}^{(4)}n_f=4d_{RR}^{(4)}n_f^2$.
These color factor identifications are also encoded in Dokshitzer's relation among the entries in Eq.~\eqref{eq:matrix} \cite{Dokshitzer:1977sg,Bukhvostov:1985vj,Bukhvostov:1985rn,Belitsky:1998gu}, valid for structures appearing for the first time at a given order of perturbation theory, which we have used to transfer available information from $\gamma_{qq}^{(3)}$, $\gamma_{qg}^{(3)}$, and $\gamma_{gq}^{(3)}$ \cite{Falcioni:2023luc,Falcioni:2023vqq,Falcioni:2024xyt,Falcioni:2024qpd,Kniehl:2025jfs,Kniehl:2025ttz,Kniehl:2026axe,Gehrmann:2026qbl} to $\gamma_{gg}^{(3)}$.

In combination with Eq.~(8) in Ref.~\cite{Falcioni:2024qpd} and the $\zeta_3$ part of Eq.~(3.14) in Ref.~\cite{Davies:2016jie}, Eqs.~\eqref{eq:quartic} and \eqref{eq:quadratic} exhaust $\gamma_{gg,\zeta_3}^{(3)}$.
For the reader's convenience, we list the complete transcendental part of $\gamma_{gg}^{(3)}$, expressed in terms of nested harmonic sums, in the Supplemental Material~\ref{supplement:2}.
Apart from the $n_f^{1,0}$ terms of $\gamma_{qg,\zeta_3}^{(3)}$ and the $n_f^{0}$ term of $\gamma_{gq,\zeta_3}^{(3)}$, all transcendental contributions to $\gamma_{ij}^{(3)}$ are now available.

Having completed our knowledge of $\gamma_{gg,\zeta_3}^{(3)}$ and conjectured specific terms of $\gamma_{gg,\mathrm{rat}}^{(3)}$, we now present the $C_F^2n_f^2$ term of the latter for all $N$.
For this particular QED-like color factor, we meanwhile avail of $\gamma_{gg}^{(3)}$ through $N=48$, considerably beyond what is published \cite{Moch:2021qrk,Moch:2023tdj,Falcioni:2024qpd,Falcioni:2025hfz}.
The values for $N=2,\ldots,44$ turn out to suffice for analytic reconstruction via the LLL algorithm \cite{Lenstra82factoringpolynomials,fplll}, leaving those for $N=46,48$ for cross checking.
We find
\begin{widetext}
\begin{eqnarray}
&&\gamma_{gg,\mathrm{rat}}^{(3)}=
%
%
\frac{32}{27}\*C_F^2\*n_f^2\*\Big\{
-\frct{169}{16}
-\frct{1234}{3}\*\D1
+\frct{1234}{3}\*\D2
+\frct{63069}{32}\*\eta
+\frct{30839}{36}\*\nu
-\frct{30813}{8}\*\D1^2
-314\*\D2^2
-\frct{267373}{48}\*\eta^2
-\frct{2330}{27}\*\nu^2
\nonumber\\&&{}
-276\*\D2^3
-\frct{15611}{3}\*\D1^3
-\frct{55291}{12}\*\eta^3
-1566\*\D1^5
-111\*\D1^6
-540\*\D1^7
-128\*\D2^5
-3162\*\eta^5
-1476\*\eta^6
-270\*\eta^7
-\frct{640}{3}\*\D2^4
\nonumber\\&&{}
-\frct{21109}{4}\*\eta^4
-\frct{29527}{12}\*\D1^4
+
4\*(45\*\eta
+90\*\eta^2
+9\*\eta^3
-15\*\nu
-8\*\nu^2)\*\HS_{-3}
+\Big(
198\*\D2
-198\*\D1
+468\*\D1^2
+72\*\D1^3
+288\*\D1^4
\nonumber\\&&{}
-32\*\D2^3
+963\*\eta^2
+684\*\eta^3
+36\*\eta^4
-\frct{32}{3}\*\D2^2
+\frct{63}{2}\*\eta
-\frct{128}{3}\*\nu^2
-\frct{859}{9}\*\nu\Big)\*\HS_{-2}
+\Big[
\frct{785}{12}\*(\D1-\D2)
+\frct{161}{12}\*\D1^2
-40\*\D1^4
+12\*\D1^5
\nonumber\\&&{}
+388\*\eta^4
+57\*\eta^5
+\frct{1099}{6}\*\D1^3
+\frct{1213}{12}\*\eta^2
-\frct{1421}{8}\*\eta
+\frct{2257}{36}\*\nu
+\frct{9353}{12}\*\eta^3\Big]\*\HS_{1}
+\Big(
109\*\eta
-323\*\D1^2
+144\*\D1^3
-126\*\D1^4
+238\*\eta^3
\nonumber\\&&{}
+66\*\eta^4
+\frct{529}{6}\*\eta^2
-\frct{239}{18}\*\nu
+\frct{583}{2}\*\D1
-\frct{583}{2}\*\D2
\Big)\*\HS_{2}
+\Big(48\*\D1
-30\*\D1^2
+36\*\D1^3
-48\*\D2
-276\*\eta
-386\*\eta^2
-96\*\eta^3
-8\*\nu^2
\nonumber\\&&{}
+\frct{244}{3}\*\nu\Big)\*\HS_{3}
+2\*(27\*\eta
+12\*\eta^2
-14\*\nu)\*\HS_{4}
-6\*(27\*\eta
+48\*\eta^2
+24\*\eta^3
-8\*\nu)\*\HS_{-2,1}
-6\*(129\*\eta
+168\*\eta^2
+36\*\eta^3
-44\*\nu)\*\HS_{1,-2}
\nonumber\\&&{}
+\Big(33\*\D1^3
+105\*\D1^4
-231\*\eta^3
-87\*\eta^4
-\frct{45}{2}\*\D1
+\frct{45}{2}\*\D2
+\frct{471}{4}\*\D1^2
+\frct{579}{2}\*\eta
-\frct{2695}{18}\*\nu
+\frct{3943}{12}\*\eta^2\Big)\*\HS_{1,1}
+\Big(
-120\*\D1
+69\*\D1^2
\nonumber\\&&{}
-102\*\D1^3
+120\*\D2
+40\*\eta^2
-42\*\eta^3
-\frct{218}{3}\*\nu
+\frct{363}{2}\*\eta\Big)\*\HS_{1,2}
+2\*(27\*\eta
+42\*\eta^2
-4\*\nu)\*\HS_{1,3}
+\Big(60\*\D1
-30\*\D1^2
+60\*\D1^3
-60\*\D2
\nonumber\\&&{}
-216\*\eta^2
-3\*\eta^3
+138\*\nu
-\frct{1377}{4}\*\eta\Big)\*\HS_{1,1,1}
+\Big(
-96\*\D1
+57\*\D1^2
-78\*\D1^3
+96\*\D2
-46\*\eta^2
-42\*\eta^3
-\frct{46}{3}\*\nu
+\frct{81}{2}\*\eta\Big)\*\HS_{2,1}
\nonumber\\&&{}
-2\*(27\*\eta
+6\*\eta^2
-16\*\nu)\*\HS_{3,1}
+12\*(3\*\eta
+3\*\eta^2
-\nu)\*(\HS_{-4}
-2\*\HS_{-3,1}
-4\*\HS_{-2,-2}
-6\*\HS_{1,-3}
-4\*\HS_{2,-2}
+4\*\HS_{1,-2,1}
+8\*\HS_{1,1,-2})
\nonumber\\&&{}
+(
-9\*\eta
-6\*\eta^2
+4\*\nu)\*(11\*\HS_{2,2}
+3\*\HS_{1,1,2}
-\HS_{1,2,1}
-7\*\HS_{2,1,1}
-11\*\HS_{1,1,1,1})\Big\}
\,.
\label{eq:rat}
\end{eqnarray}
\end{widetext}
Equations~\eqref{eq:quartic}, \eqref{eq:quadratic}, and \eqref{eq:rat} represent our key results and mark an important step towards establishing exact $N$ dependencies of the $\gamma_{ij}^{(3)}$ set governing DGLAP evolution of PDFs at N${}^3$LO.

We briefly review our workflow.
To be specific, we consider $\gamma_{gg,\zeta_3}^{(3)}$.
We have $\gamma_{\mathrm{ns},\zeta_3}^{(3)+}$ \cite{Kniehl:2025jfs,Kniehl:2026axe,Gehrmann:2026qbl} and $\gamma_{\mathrm{ps},\zeta_3}^{(3)}$ \cite{Falcioni:2025hfz} for all $N$, and $\gamma_{gg,\zeta_3}^{(3)}$ \cite{Moch:2021qrk,Moch:2023tdj,Falcioni:2024qpd,Falcioni:2025hfz} and hence also $\tr\hat{\gamma}_{\zeta_3}^{(3)}=\gamma_{\mathrm{ns},\zeta_3}^{(3)+}+\gamma_{\mathrm{ps},\zeta_3}^{(3)}+\gamma_{gg,\zeta_3}^{(3)}$ for a set of low-$N$ values.
We evaluate the non-reciprocity-respecting component $\NP_{\zeta_3}^{(3)}$ of $\tr\hat{\gamma}_{\zeta_3}^{(3)}$ for all $N$ from sub-N${}^3$LO results via Eq.~(14) of Ref.~\cite{Kniehl:2026axe}.
We write $\mathcal{P}_{\zeta_3}^{(3)}=\tr\hat{\gamma}_{\zeta_3}^{(3)}+\NP_{\zeta_3}^{(3)}$ as a linear combination, with rational coefficients, of all binomial harmonic sums, bare or dressed by powers of $\eta$ or $\nu$, of overall weight $w\le4$.
We reduce the number of coefficients in this ansatz by imposing constraints from $\mathcal{N}=1,4$ SYM via Dokshitzer's relation \cite{Dokshitzer:1977sg,Bukhvostov:1985vj,Bukhvostov:1985rn,Belitsky:1998gu} and maximal-transcendentality principle \cite{Kotikov:2002ab,Kotikov:2006ts}, respectively.
Matching the low-$N$ values of $\mathcal{P}_{\zeta_3}^{(3)}$, we fix the coefficients of $\eta^{4,3,2}$ and $\nu^4$ using the Chinese remainder theorem \cite{china} and analytically reconstruct the remaining ones with \texttt{fplll} \cite{fplll}.
We end up with $\gamma_{gg,\zeta_3}^{(3)}=\mathcal{P}_{\zeta_3}^{(3)}-\NP_{\zeta_3}^{(3)}-\gamma_{\mathrm{ns},\zeta_3}^{(3)+}-\gamma_{\mathrm{ps},\zeta_3}^{(3)}$ for all $N$.
The evaluation of the $C_F^2n_f^2$ term of $\gamma_{gg,\mathrm{rat}}^{(3)}$ proceeds similarly.

We conclude with an outlook.
The present supply of analytic low-$N$ $\gamma_{ij}^{(3)}$ values, obtained with the FORCER program~\cite{Ruijl:2017cxj} written in FORM~\cite{Vermaseren:2000nd,Kuipers:2012rf,Ruijl:2017dtg}, needs further extension to enable analytic reconstruction of the as-yet missing all-$N$ expressions.
While this appears feasible for contributions involving $\zeta_3$ and/or $n_f$, non-fermionic rational contributions with $w=7$ may prove more difficult, as they require considerably more input.
Further progress may come from the alternative approach recently developed and applied to the QED-like $C_F^3n_f$ contributions to $\gamma_{\mathrm{ns}}^{(3)\pm}$ in Ref.~\cite{Gehrmann:2023iah} and to $\gamma_{\mathrm{ns},\mathrm{rat}}^{(3)\pm}$ and
$\gamma_{\mathrm{ns},\mathrm{rat}}^{(3)\mathrm{s}}$ in Ref.~\cite{Gehrmann:2026qbl}.
This relies on the computation of off-shell operator matrix elements and employs integration-by-parts reductions and differential equations with respect to a tracing parameter.
It is reassuring to notice that our recent LLL-based \cite{Lenstra82factoringpolynomials,fplll} analytic reconstructions of transcendental \cite{Kniehl:2025jfs,Kniehl:2026axe} and fermionic \cite{Kniehl:2025ttz} contributions to $\gamma_{\mathrm{ns}}^{(3)}$ were confirmed by Ref.~\cite{Gehrmann:2026qbl}.
The study of scaling violations in PDFs remains a challenging and intriguing topic of considerable phenomenological importance and broad interest, as the progress presented here exploits symmetries and structural relations in quantum field theory beyond QCD, also linking developments across different areas of theoretical physics.

\vspace{2mm}
\begin{acknowledgments}
\emph{Acknowledgments:}  
B.A.K. and V.N.V. were supported in part by Deutsche Forschungsgemeinschaft (DFG) through Grants KN~365/13-2 and KN~365/16-1, S.O.M. by European Research Council (ERC) through Advanced Grant 101095857 {\it Conformal-EIC}, and A.V. by UK Science and Technology Facilities Council (STFC) through Consolidated Grant ST/X000699/1 and by Alexander von Humboldt-Stiftung (AvH) through a Research Award.
\end{acknowledgments}





\newpage
\cleardoublepage
\appendix
%
%
\renewcommand{\theequation}{A.\arabic{equation}}
\setcounter{equation}{0}
\renewcommand{\thefigure}{A.\arabic{figure}}
\setcounter{figure}{0}
\renewcommand{\thetable}{A.\arabic{table}}
\setcounter{table}{0}

\begin{widetext}
\setcounter{secnumdepth}{2} 
\section{Supplemental Material}

\subsection{Typical application of \texttt{\textbf{fplll}}}
\label{supplement:1}

As a typical example for the analytic reconstruction using the program \texttt{fplll} \cite{fplll}, we consider the coefficient of $C_F^3\nf$ in $\gamma^{(3)}_{gg,\zeta_3}$.
We write down the full basis of binomial harmonic sums, bare and dressed by powers of $\eta$ and $\nu$, with overall weights $w\le4$,
\begin{eqnarray}
&&\{1, \eta, \nu, \eta^2, \nu^2, \BS_{1}, \eta \BS_{1}, 
 \nu \BS_{1}, \eta^3, \nu^3, \eta^2 \BS_{1}, \nu^2 \BS_{1}, \BS_{2}, 
 \eta \BS_{2}, \nu \BS_{2}, \BS_{1, 1}, \eta \BS_{1, 1}, 
 \nu \BS_{1, 1}, \eta^4, \nu^4, \eta^3 \BS_{1}, \nu^3 \BS_{1}, 
 \eta^2 \BS_{2}, \nu^2 \BS_{2}, \BS_{3}, \nonumber\\&&\ \ \
 \eta \BS_{3}, \nu \BS_{3}, \eta^2 \BS_{1, 1}, \nu^2 \BS_{1, 1}, \BS_{1, 2}, \eta \BS_{1, 2}, 
 \nu \BS_{1, 2}, \BS_{2, 1}, \eta \BS_{2, 1}, \nu \BS_{2, 1}\}\,.
 \label{BasisCF3nf}
\end{eqnarray}
It is useful to rescale the basis elements in Eq.~\eqref{BasisCF3nf} by a common factor $2^i/3^j$ to make their coefficients come out as small numbers.
Here, $32/3$ is found to yield optimal results.

We use as inputs the nine even Mellin moments with $N=4,6,\ldots,20$ \cite{Moch:2021qrk,Moch:2023tdj,Falcioni:2024qpd}, leaving the one with $N=22$ \cite{Falcioni:2025hfz} for cross checking.
We refrain from inputting the value for $N=2$ \cite{Moch:2021qrk}, bearing in mind that it might be contaminated by $\delta_{N,2}$, similarly to the $C_AC_Fn_f^2$ contribution to $\gamma_{gg,\zeta_3}^{(3)}$ \cite{Falcioni:2024qpd}.
Applying \texttt{fplll} with parameters ``\texttt{-f mpfr -p 100 -a bkz -b 10}'' to the matrix thus constructed, we find the first non-trivial vector of coefficients to be
\begin{eqnarray}
&&\{-1334,-860,1081,908,-504,95,-54,-359,28,0,188,272,225,1013,-280,0,-985,332,168,0,-1848,0,696,
\nonumber\\&&\ \  160,-712,-756,466,-240,-160,-93,408,-58,785,348,-408,-2\}\,,
\label{eq:one}
\end{eqnarray}
where the last entry is the normalization factor.
Eq.~\eqref{eq:one} is far to complicated to be true.

As is well known from previous results, the coefficients of the first few basis elements in Eq.~\eqref{BasisCF3nf}, namely, the lower powers of $\eta$ or $\nu$ and the simplest binomial harmonic sums like $\BS_{1}$, are relatively ``complicated'' numbers upon factoring out $2^i3^j$.
Therefore, the chances of \texttt{fplll} to succeed may be increased by eliminating one of them using one out of the nine available equations.
Eliminating the first basis element, $1$, using the first equation, for $N=4$, we obtain
\begin{eqnarray}
&&\{-889,199,-1816,1275,37,-419,-107,-56,-280,-275,-210,107,-194,-366,0,-568,181,-336,96,6,-24,\nonumber\\&&\ \ 
  -348,108,7,709,-204,262,-132,48,-23,178,-158,1030,26,4\}\,,
\label{eq:two}
\end{eqnarray}
which does not qualify.
Eliminating, one after another, also $\eta,\nu,\eta^2$ using, in this order, the equations for $N=6,8,10$, we get
\begin{eqnarray}
  &&\{-731,908,-264,-132,601,100,28,24,-169,106,-473,-638,176,0,-449,154,168,0,-246,0,954,80,255,
  \nonumber\\&&\ \ 
  177,80,-84,-80,45,177,-200,-320,-354,120,-2\}\,,\nonumber\\
&&\{454,30,-205,88,-364,14,84,178,-129,-11,-493,-372,0,219,-73,84,48,176,-12,364,-136,-13,48,\nonumber\\&&\ \ 
  -306,-232,124,107,240,126,-104,-288,180,-1\}\,,\nonumber\\
 &&\{85,-205,-254,95,14,136,19,-144,180,135,-153,0,123,-41,84,-36,-108,9,51,-88,-113,21,91,\nonumber\\&&\ \ 
  -177,97,107,-75,1,-4,54,-92,-1\}\,,
\label{eq:three}
\end{eqnarray}
none of which qualifies either.

However, if we trade $\nu$ by $\BS_{1}$, we find 
\begin{eqnarray}
&&\{-27,48,219,-84,14,0,354,-24,30,132,-56,0,-66,24,84,0,192,0,216,0,-12,0,0,-60,0,-12,0,0,\nonumber\\&&\ \ 
14,0,0,-1\}\,,
\label{eq:four}
\end{eqnarray}
which strikes as the solution.
In fact, the vector in Eq.~\eqref{eq:four} distinctly differs from those in Eqs.~\eqref{eq:one}--\eqref{eq:three}, by carrying numerous zeros alongside integers that can be factored as $2^i3^j$.
This qualitative breakthrough serves as the primary argument in favor of Eq.~\eqref{eq:four}.

In the case at hand, we still have the equation for $N=22$ \cite{Falcioni:2025hfz} and the Chinese remainder theorem \cite{china}, fixing the coefficients of $\eta^{4,3,2}$ and $\nu^4$ \cite{Moch:2014sna}, for cross checks, which all work out all right.
In other cases, when all available information is just enough to produce a solution with such distinctive properties as in Eq.~\eqref{eq:four}, we are still on the safe side.

\boldmath
\subsection{All-$N$ results for $\gamma_{gg,\zeta_k}^{(3)}$ with $k=3,4,5$}
\label{supplement:2}
\unboldmath

For the reader's convenience, we list here the complete all-$N$ expressions for the transcendental contributions $\gamma_{gg,\zeta_k}^{(3)}$ with $k=3,4,5$ in
Eq.~\eqref{eq:zeta3}:

\begin{eqnarray}
&&\gamma_{gg,\zeta_3}^{(3)}=
%
%
\frct{32}{3}\*C_F^3\*n_f\*
\Big[
24\*\D1
-24\*\D2
+231\*\D1^2
+294\*\D1^3
+144\*\D1^4
-\frct{897}{4}\*\eta
-16\*\nu
-177\*\eta^2
-138\*\eta^3
-114\*\eta^4
+6\*(
6\*\D1
\nonumber\\
&&{}
-6\*\D2
-2\*\D1^2
+8\*\D1^3
-71\*\eta
+20\*\nu
-78\*\eta^2
-56\*\eta^3
)\*\HS_{1}
-4\*(15\*\eta
-4\*\nu
+18\*\eta^2
)\*\HS_{2}
+4\*(33\*\eta
-16\*\nu
+90\*\eta^2
)\*\HS_{-2}
+12\*(13\*\eta
\nonumber\\
&&{}-4\*\nu
+14\*\eta^2
)\*\HS_{1,1}
-\frct{2}{3}\*\delta_{N,2}
\Big]
+\frct{32}{9}\*C_A\*C_F^2\*n_f\*\Big\{
-\frct{11}{2}
+\frct{1161}{2}\*(\D1-\D2)
-2817\*\D1^2
-1476\*\D1^3
-1836\*\D1^4
+148\*\D2^2
\nonumber\\
&&{}
+120\*\D2^3
+176\*\nu^2
+\frct{6879}{2}\*\eta
-\frct{1597}{3}\*\nu
+\frct{3771}{2}\*\eta^2
+2376\*\eta^3
+1224\*\eta^4
+3\*\Big[\frct{37}{2}
-30\*(\D1-\D2)
+123\*\D1^2
+12\*\D2^2
+132\*\D1^3
\nonumber\\
&&{}
+1065\*\eta
-366\*\nu
+1656\*\eta^2
-8\*\nu^2
+930\*\eta^3
\Big]\*\HS_{1}
+3\*(231\*\eta
-68\*\nu
+330\*\eta^2
)\*\HS_{2}
-24\*(24\*\eta
-13\*\nu
+81\*\eta^2
)\*\HS_{-2}
-48\*(30\*\eta
-10\*\nu
\nonumber\\
&&{}
+39\*\eta^2
)\*\HS_{1,1}
+\frct{7}{3}\*\delta_{N,2}
\Big\}
+\frct{32}{3}\*C_A^2\*C_F\*n_f\*\Big\{\frct{2}{3}
-\frct{471}{2}\*(\D1-\D2)
+\frct{2019}{2}\*\D1^2
-\frct{514}{3}\*\D2^2
+201\*\D1^3
-136\*\D2^3
+684\*\D1^4
\nonumber\\
&&{}
-\frct{2530}{3}\*\eta
+\frct{1303}{18}\*\nu
-\frct{1325}{3}\*\eta^2
-\frct{256}{3}\*\nu^2
-892\*\eta^3
-40\*\nu^3
-420\*\eta^4
+\Big[\frct{241}{6}
-6\*(\D1-\D2)
-135\*\D1^2
-84\*\D2^2
-324\*\D1^3
\nonumber\\
&&{}
-552\*\eta
+\frct{773}{3}\*\nu
-1470\*\eta^2
+112\*\nu^2
-726\*\eta^3
\Big]\*\HS_{1}
-(
309\*\eta
-124\*\nu
+426\*\eta^2
)\*\HS_{2}
+2\*(3
-3\*\eta
-8\*\nu
+210\*\eta^2
)\*\HS_{-2}
+8\*(60\*\eta
\nonumber\\
&&{}
-23\*\nu
+90\*\eta^2
)\*\HS_{1,1}
+8\*\HS_{3}
+8\*\HS_{-3}
-24\*\HS_{1,2}
-24\*\HS_{1,-2}
-24\*\HS_{2,1}
+8\*\HS_{-2,1}
-\frct{2}{9}\*\delta_{N,2}
\Big\}
+\frct{32}{3}\*C_A^3\*n_f\*\Big[
\frct{2197}{360}
-294\*\D1^2
\nonumber\\
&&{}
+122\*\D2^2
-24\*\D1^3
+96\*\D2^3
-216\*\D1^4
+\frct{683}{12}\*\eta
+\frct{6734}{135}\*\nu
+\frct{2431}{12}\*\eta^2
+\frct{194}{3}\*\nu^2
+377\*\eta^3
+20\*\nu^3
+144\*\eta^4
+\frct{1}{18}\*(
-1173
+432\*\D1^2
\nonumber\\
&&{}
+1296\*\D2^2
+2592\*\D1^3
+4989\*\eta
-1696\*\nu
+8913\*\eta^2
-1440\*\nu^2
+4194\*\eta^3
)\*\HS_{1}
+\frct{5}{3}\*(99\*\eta
-40\*\nu
+54\*\eta^2
)\*\HS_{2}
-\frct{2}{3}\*(
315\*\eta
-92\*\nu
\nonumber\\
&&{}
+162\*\eta^2
)\*\HS_{1,1}
+\frct{1}{6}\*\Big(
\frct{128}{15}
+\frct{2436}{5}\*\eta
-116\*\nu
-1056\*\eta^2
\Big)\*\HS_{-2}
-\frct{1}{6}\*(
59
-18\*\eta)\*\HS_{3}
+\frct{1}{3}\*(31
-18\*\eta)\*\HS_{-3}
+24\*\HS_{1,2}
+\frct{46}{3}\*\HS_{1,-2}
\nonumber\\
&&{}
+24\*\HS_{2,1}
+12\*(
-3
+\eta)\*\HS_{-2,1}
+\frct{1}{2}\*N\*(N+1)\*\Big(
\frct{4}{45}\*\HS_{-2}
+\HS_{3}
-2\*\HS_{-3}
+4\*\HS_{-2,1}\Big)\Big]
+\frct{64}{3}\*C_A^4\*\Big[
-\frct{2621}{720}
-\frct{36013}{144}\*\eta
+\frct{101027}{1080}\*\nu
\nonumber\\
&&{}
-\frct{1133}{6}\*\eta^2
-\frct{207}{4}\*\nu^2
-97\*\eta^3
+19\*\nu^3
-12\*\eta^4
-24\*\nu^4
+\frct{1}{36}\*(437
-21021\*\eta
+5915\*\nu
-12174\*\eta^2
-2664\*\nu^2
-4320\*\eta^3
\nonumber\\
&&{}
+1728\*\nu^3)\*\HS_{1}
+\frct{1}{3}\*(5
-51\*\eta
+33\*\nu)\*\HS_{3}
+6\*\HS_{4}
+\frct{4}{3}\*(45\*\eta
+45\*\eta^2
-25\*\nu
+12\*\nu^2)\*(\HS_{2}-2\*\HS_{1,1})
-\frct{1}{18}\*\Big(
\frct{452}{5}
-\frct{1023}{5}\*\eta
+435\*\nu
\nonumber\\
&&{}
-612\*\eta^2
+72\*\nu^2
\Big)\*\HS_{-2}
-\frct{1}{3}\*(
131
-141\*\eta
+105\*\nu)\*\HS_{-3}
-18\*\HS_{-4}
-11\*\HS_{1,3}
+\frct{1}{6}\*(143
-96\*\eta
+96\*\nu)\*\HS_{1,-2}
+(127
-156\*\eta
\nonumber\\
&&{}
+108\*\nu)\*\frct{1}{2}\*\HS_{-2,1}
+35\*\HS_{1,-3}
+8\*\HS_{2,-2}
-5\*\HS_{3,1}
+4\*\HS_{-2,-2}
+16\*\HS_{-2,2}
+43\*\HS_{-3,1}
-16\*\HS_{1,1,-2}
-54\*\HS_{1,-2,1}
-32\*\HS_{-2,1,1}
\nonumber\\
&&{}
-N\*(N+1)\*\Big(
\frct{1}{90}\*\HS_{-2}
+\frct{1}{4}\*\HS_{3}
-\frct{1}{2}\*\HS_{-3}
+\HS_{-2,1}
\Big)
\Big]
+\frct{64}{3}\*\frct{d_{AA}^{(4)}}{n_a}\*
  \Big[  
 \frct{331}{15} 
+ \frct{429}{2}\*\eta
- \frct{3824}{45}\*\nu
+ 180\*\eta^2 
+ 2\*\HS_{1} \*\Big(-5 + 1352\*\eta - \frct{1328}{3}\*\nu 
\nonumber\\ 
&&{}
+ 722\*\eta^2 
+ 192\*\nu^2
+ 144\*\eta^3 
\Big)
- 4\*\HS_{-2}\*\Big(\frct{1126}{15}+\frct{1127}{5}\*\eta 
- 19\*\nu + 48\*\nu^2 - 12\*\eta^2 \Big)
- 16\*(2\*\HS_{1,1}-\HS_{2})\*(63\*\eta - 25\*\nu + 54\*\eta^2 +12\*\nu^2)
\nonumber\\ 
&&{}
+4\*(31-102\*\eta+30\*\nu)\*\HS_{-3}
-4\*(43-66\*\eta+30\*\nu)\*\HS_{3}
-36\*(13-28\*\eta+12\*\nu)\*\HS_{-2,1}
+4\*(55-48\*\eta+48\*\nu)\*\HS_{1,-2}
-24\*\HS_{-3,1}
\nonumber\\ 
&&{}
-96\*\HS_{-2,-2}
+192\*\HS_{-2,2}
-120\*\HS_{1,-3}
+120\*\HS_{1,3}
+96\*\HS_{2,-2}
-24\*\HS_{3,1}
-384\*\HS_{-2,1,1}
+432\*\HS_{1,-2,1}
-192\*\HS_{1,1,-2}
\nonumber\\ 
&&{}
-2\* N\*(N+1)\*
   \Big(
\frct{1}{15}\* \HS_{-2}
+ 6\* \HS_{-3}
- 3\* \HS_{3}
- 12\* \HS_{-2, 1}
\Big) 
\Big]
%
+\frct{128}{3}\*\frct{d_{RA}^{(4)}}{n_a}\*n_f\*\Big[  
\frct{103}{15}
+\frct{2908}{45}\*\nu
-291\*\eta 
-387\*\eta^2
+2\*\Big(
5 
+ 14\*\eta 
+\frct{52}{3}\*\nu 
\nonumber\\ 
&&{}
- 46\*\eta ^2 - 84\*\eta ^3\Big)\*\HS_{1}
+2\*\Big(\frct{514}{15}+\frct{618}{5}\*\eta
- 4\*\nu 
- 132\*\eta^2\Big)\*\HS_{-2}
+2\*\big(9\*\eta - 16\*\nu + 198\*\eta^2\big)\*(2\*\HS_{1,1}-\HS_{2})
- 4\* (13 - 18\* \eta)\* \HS_{-3}
\nonumber\\
&&{}
+ 2\* (23 - 18\* \eta)\* \HS_{3}
+ 144\* (1 - \eta)\* \HS_{-2, 1}
- 40\* \HS_{1, -2}
+(N+1)\*N\*\Big(
\frct{4}{15}\* \HS_{-2}
+ 12\* \HS_{-3}
- 6\* \HS_{3}
- 24\* \HS_{-2, 1}
\Big)
\Big]
+\frct{16}{27}\*C_A^2\*n_f^2\*\Big\{
-\frct{13}{10}
\nonumber\\
&&{}
+\frct{4773}{4}\*\eta
-\frct{7154}{15}\*\nu
+\frct{2577}{2}\*\eta^2
-84\*\nu^2
+324\*\eta^3
-\frct{2}{5}\*
[
N\*(N+1)
+17
+42\*\eta
]\*\HS_{-2}
+
(120
-753\*\eta
+230\*\nu
-492\*\eta^2
+144\*\eta^3
)\*\HS_{1}
\nonumber\\
&&{}
-36\*\eta\*
(
1
-4\*\eta)\*\HS_{2}
+72\*\eta\*
(
1
-4\*\eta)\*\HS_{1,1}\Big\}
+\frct{256}{9}\*\frct{d_{RR}^{(4)}}{n_a}\*n_f^2\*\Big\{
-\frct{29}{5}
+\frct{297}{2}\*\eta
-\frct{299}{15}\*\nu
+135\*\eta^2
-\frct{2}{5}\*
[17
+42\*\eta
+N\*(N+1) ] \*\HS_{-2}
\nonumber\\
&&{}
-2\*
(
57\*\eta
-\nu
-78\*\eta^2
-72\*\eta^3
)\*\HS_{1}
-36\*\eta\*
(
1
-4\*\eta)\*\HS_{2}
+72\*\eta\*
(
1
-4\*\eta)\*\HS_{1,1}\Big\}
+\frct{32}{9}\*C_A\*C_F\*n_f^2\*\Big[
-5
-21\*(\D1-\D2)
-\frct{799}{2}\*\eta
\nonumber\\
&&{}
+143\*\nu
+15\*\D1^2
-588\*\eta^2
+24\*\nu^2
-12\*\D1^3
-150\*\eta^3
-4\*
(
5
-54\*\eta
+19\*\nu
-60\*\eta^2
)\*\HS_{1}
+\frct{1}{3}\*\delta_{N,2}
\Big]
+\frct{32}{9}\*C_F^2\*n_f^2\*\Big[
48\*(\D1-\D2)
\nonumber\\
&&{}
+\frct{11}{2}
+231\*\eta
-\frct{214}{3}\*\nu
-24\*\D1^2
+395\*\eta^2
-8\*\nu^2
+48\*\D1^3
+90\*\eta^3
-2\*
(63\*\eta
-20\*\nu
+66\*\eta^2
)\*\HS_{1}
\Big]
+
\frct{64}{27}\*C_A\*n_f^3\*
(\eta
-\nu
+\HS_{1})
\nonumber\\
&&{}
+\frct{32}{9}\*C_F\*n_f^3\*
(9\*\eta
-4\*\nu
+6\*\eta^2)
\,,
\\
&&\gamma_{gg,\zeta_4}^{(3)}=
%
%
176\*C_A^3\*n_f\*
(4\*\eta
-2\*\nu
+3\*\eta^2
+\HS_{1})
-8\*C_A^2\*C_F\*n_f\*
[25\*\eta
-12\*\nu
+60\*\eta^2
-16\*\nu^2
-24\*\eta^3
+2\*
(11
-18\*\eta
+8\*\nu
-12\*\eta^2
)\*\HS_{1}]
\nonumber\\
&&{}
-16\*C_A\*C_F^2\*n_f\*
[12\*\eta
-8\*\nu
-15\*\eta^2
+8\*\nu^2
+6\*\eta^3
+4\*
(9\*\eta
-4\*\nu
+6\*\eta^2
)\*\HS_{1}]
-8\*C_F^3\*n_f\*
[39\*\eta
-16\*\nu
+36\*\eta^2
+12\*\eta^3
-4\*
(9\*\eta
-4\*\nu
\nonumber\\
&&{}
+6\*\eta^2
)\*\HS_{1}]
-32\*C_A^2\*n_f^2\*
(4\*\eta
-2\*\nu
+3\*\eta^2
+\HS_{1})
+\frct{16}{3}\*C_A\*C_F\*n_f^2\*
(15\*\eta
-4\*\nu
+24\*\eta^2
+6\*\HS_{1})
+\frct{16}{3}\*C_F^2\*n_f^2\*
(9\*\eta
-8\*\nu
-6\*\eta^2
   )
\,,
\\
%
&&\gamma_{gg,\zeta_5}^{(3)}=
%
%
\frct{160}{9}\*\Big(C_A^4+12\*\frct{d_{AA}^{(4)}}{n_a}\Big)\*
\Big[
-\frct{1}{60}\*N\*(N+1)
-\frct{751}{30}
-21\*\nu
+\frct{94}{5}\*\eta
-12\*\nu^2
+36\*\eta^2
+(33
+24\*\nu
-24\*\eta)\*\HS_{1}
+12\*\HS_{2}
\nonumber\\ &&{}
-24\*\HS_{1,1}
\Big]
+\frct{80}{9}\*C_A^3\*n_f\*
\Big[
\frct{1}{15}\*N\*(N+1)
+\frct{17}{15}
+16\*\nu
+\frct{149}{5}\*\eta
-456\*\eta^2
+6\*\HS_{1}
\Big]
+\frct{640}{3}\*\frct{d_{RA}^{(4)}}{n_a}\*n_f\*\Big[
\frct{287}{15}
+\frct{1}{15}\*N\*(N+1)
+4\*\nu
\nonumber\\ &&{}
-\frct{121}{5}\*\eta
+30\*\eta^2
-12\*\HS_{1}
\Big]
+80\*C_A^2\*C_F\*n_f\*\Big(2
-\frct{68}{3}\*\nu
+41\*\eta
+184\*\eta^2
+2\*\HS_{1}
\Big)
+80\*C_A\*C_F^2\*n_f\*\big(32\*\nu
-81\*\eta
-262\*\eta^2
-4\*\HS_{1}
\big)
\nonumber\\ &&{}
+160\*C_F^3\*n_f\*\Big(
-\frct{16}{3}\*\nu
+17\*\eta
+66\*\eta^2
\Big)
%
+\frct{8}{9}\*\Big(C_A^2+48\*\frct{d_{RR}^{(4)}}{n_a}\Big)\*n_f^2\*\Big[
-\frct{1}{3}\*N\*(N+1)
-\frct{17}{3}
+46\*\eta
-240\*\eta^2
\Big]
\,,
\end{eqnarray}
where the notation is as in the main text.
\end{widetext}

\end{document}